# BARIONIC RESONANCES WITH POSITIVE STRANGENESS S=+1 IN THE SYSTEM $nK^+$ FROM THE REACTION $np \to npK^+K^-$ AT THE MOMENTUM OF INCIDENT NEUTRONS $P_n$=5.20 GeV/c

Yu.A.Troyan[1*], A.V.Beljaev[1], A.Yu.Troyan[1], E.B.Plekhanov[1], A.P.Jerusalimov[1], S.G.Arakelian[2]

[1] - Veksler and Baldin Laboratory of High Energies, JINR, Dubna.
[2] - Lebedev Institute of Physics, Russian Academy of Sciences, Moscow

*-E-mail: atroyan@jinr.ru

## Abstract

The production and properties of the resonances with the strangeness S = +1 in the system of $nK^+$ were studied in the reaction $np \to npK^+K^-$ at the momentum of incident neutrons $P_n$ = (5.20 ± 0.12) GeV/c. A number of peculiarities were found in the effective mass spectrum of the mentioned above system. Their widths are comparable with the mass resolution. The estimation of the spins of resonances was carried out and the rotational band connecting the resonances masses and their spins was constructed.

## 1.    Introduction

The extension of a SU(3) symmetry scheme for states with the positive strangeness S = +1 has been suggested by Diakonov, Petrov, and Polyakov [1, 2]. They claimed existence of antidecuplet $\overline{10}$, which includes states that contain five quarks *(uudd$\overline{s}$)*. The dynamics of new resonances were based on the chiral soliton model.

This fact made it possible to estimate the masses, widths and quantum numbers of expected new states, to propose a formula of the rotational band that relates the resonance masses to their spins. The $\Theta$ resonance of mass M=1530 MeV/c$^2$, width $\Gamma \leq 15$ MeV/c$^2$, hypercharge Y = 2, isotopic spin I = 0, and spin-parity $J^P = 1/2^+$ is on the top of the antidecuplet.

This hypothesis is discussed in detail in many theoretical studies [3].

There are quite different approaches to the problem of these resonances, for example:

- composition of quarks into diquarks, which is accompanied by a rise of super-conducting layers [4];
- a pure quark picture, where isoscalar, isovector and isotensor states consisting of five quarks (both usual and strange) can arise, which greatly extends possible quantum numbers and resonance masses as well as the probability of resonance decays (for example, the $\Theta$ resonances may have the quantum numbers $1/2^-$, $3/2^-$, $5/2^-$) [5].

The properties of the particles from the predicted antidecuplet allow a direct search for effects. These are both moderate masses and widths accessible for direct measurements.

This stimulated a number of experimental works [3] in which a resonance with a mass of ~ 1540 MeV/c$^2$ and width from 3 to 25 MeV/c$^2$ was discovered in the $nK^+$ or $pK_s^0$ systems.

None of these works has observed the rotational band or more than one resonance state in the $nK^+$ system. More than one resonance state was observed only in the work by P. Aslanyan [6] in the $pK_s^0$ system.

Also, neither spin of resonance nor its parity was determined.

And, some works declared the absence of effects in the above systems.

In present work we have attempted to study in greater detail the characteristics of effects observed by us.

This work is continuation of our investigation [7] of the $nK^+$ - system with increased statistic.

## 2. Experiment

The study was carried out using the data obtained in an exposure of 1-*m* HBC of LHE (JINR) to a quasimonochromatic neutron beam with $\Delta P_n/P_n \approx 2.5\%$, $\Delta\Omega_n \approx 10^{-7}$ *sterad*. due to the acceleration of deuterons by synchrophasotron of LHE. The neutron channel has been detailed in [8]

The accuracy of momenta of secondary charged particles from the reaction $np \to npK^+K^-$ are: $\delta P \approx 2\%$ for protons and $\delta P \approx 3\%$ for $K^+$ and $K^-$. The angular accuracy was $\leq 0.5°$.

The channels of the reactions were separated by the standard $\chi^2$ – method taking into account the corresponding coupling equations [9-11].

There is only one coupling equation for the parameters of the reaction $np \to npK^+K^-$ (energy conservation law) and the experimental $\chi^2$ – distribution must be the same as the theoretical $\chi^2$ – distribution with one degree of freedom.

The experimental (histogram) and theoretical (curve) $\chi^2$ distributions for the above reaction are shown in figure 1a.

One can see a good agreement between them up to $\chi^2 = 3$ and some difference for $\chi^2 > 3$.

Figure 1b shows the missing mass distribution for the $\chi^2 \leq 3$ events. One can see that the distribution has a maximum at the missing mass equal to the neutron mass with accuracy of 0.1 MeV/c$^2$ and is symmetric about the neutron mass. The width at the half-height is 7 MeV/c$^2$.

For more purity of data a small number of events with the missing masses out of green range were excluded.

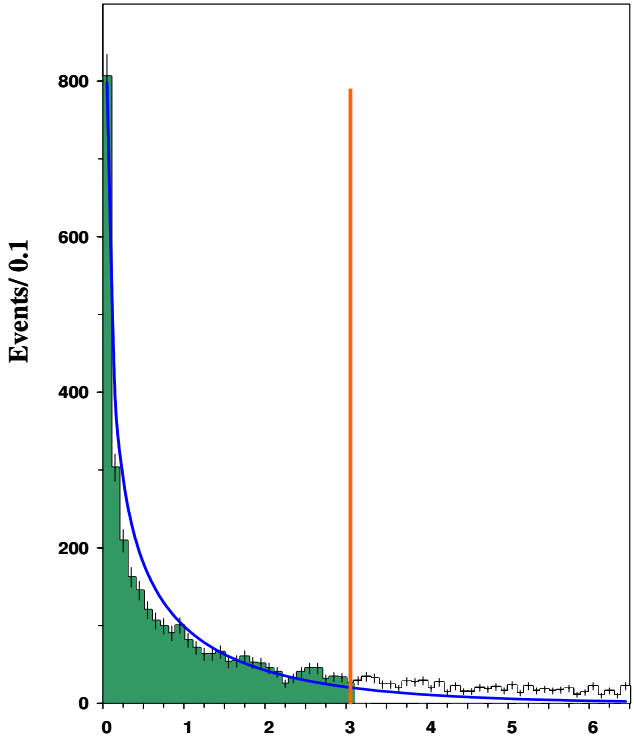
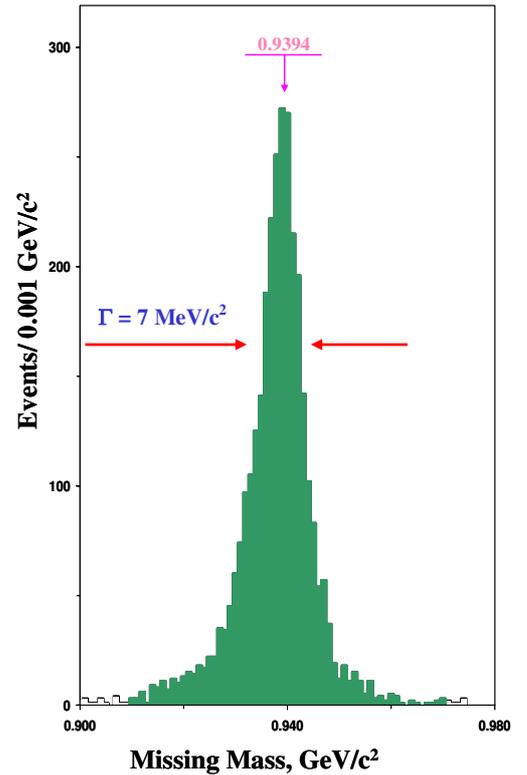

Figure 1a
The experimental (histogram) and the theoretical (curve) $\chi^2$ –distributions for the reaction $np \to npK^+K^-$.

Figure 1b
The missing mass distributions for the reaction $np \to npK^+K^-$.

As a result, 3118 events were selected under the 4π geometry (the absence of any angular selections)

## 3. Effective mass distribution

Figure 2 shows the effective mass distribution of $nK^+$- combinations from the reaction $np \to npK^+K^-$ at $P_n$ = (5.20 ± 0.12) GeV/c.

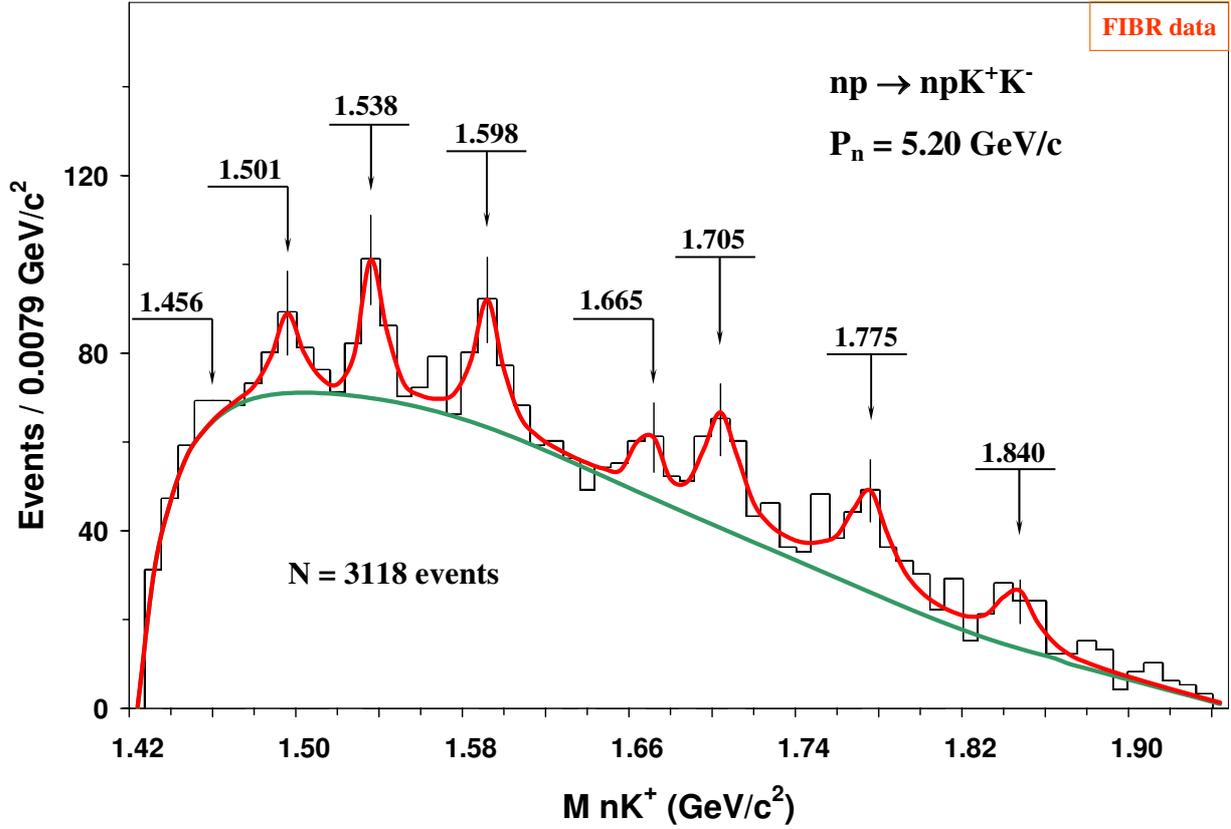

Figure.2

The effective mass distribution of $nK^+$- combinations from the reaction $np \to npK^+K^-$ at $P_n$ = (5.20 ± 0.12) GeV/c.
Green line – the background curve taken in the form of Legendre polynomial of $9^{th}$ degree.
Red line – the sum of the background curve and the 7 resonance curves taken in the Breight-Wigner form.

The mass distribution is approximated by an incoherent sum of the background curve taken in the form of a superposition of Legendre polynomials and by resonance curves taken in the Breight-Wigner form.

The requirements to the background curve are the following:
- firstly, the errors of the coefficients must be not more than 50 % for each term of the polynomial;
- secondly, the polynomial must describe the experimental distribution after "deletion" of resonance regions with $\overline{\chi^2} = 1.0$ and $\sqrt{D} = 1.41$ (the parameters of $\chi^2$ – distribution with 1 degree of freedom).

The distribution on Fig.2 is approximated by Legendre polynomial of the $9^{th}$ power and by seven resonance curves. The experimental values of the resonance masses (obtained by fitting procedure) are shown by arrows. The values $\overline{\chi^2}$ and $\sqrt{D}$ for the background curve of the distribution in Fig.2 (with resonance regions excluded) are equal to $\overline{\chi^2} = 0.86 \pm 0.24$ and $\sqrt{D} = 1.28 \pm 0.17$. The contribution of the background to this distribution is 84%. The same values for the background curve normalized to 100% of events in the plot (with resonance regions included) are equal to $\overline{\chi^2} = 1.55 \pm 0.18$ and $\sqrt{D} = 1.90 \pm 0.13$ (Confidence Level 0.3%).

It is valid to note that the examination of mass region ≈1500 MeV/c² with small step give us two very narrow peaks at the masses 1498 MeV/c² and 1507 MeV/c² which result in the average value of an experimental mass equal to M=1501 MeV/c² for the mass distribution shown above.

The results of approximation are presented in Table 1.

Table 1

|   | $M_{Res} \pm \Delta M_{Res}$, MeV/c² | $\Gamma_{Res}^{exp} \pm \Delta \Gamma_{Res}^{exp}$, MeV/c² | S.D. | σ μb |
|---|---|---|---|---|
| 1 | 1501 ± 5 | 13 ± 8 | 2.6 | 6 ± 4 |
| 2 | 1538 ± 3 | 9 ± 4 | 4.1 | 10 ± 4 |
| 3 | 1598 ± 4 | 13 ± 5 | 4.3 | 11 ± 4 |
| 4 | 1665 ± 4 | 8 ± 7 | 2.6 | 5 ± 3 |
| 5 | 1705 ± 4 | 19 ± 6 | 5.8 | 11 ± 3 |
| 6 | 1775 ± 4 | 29 ± 9 | 6.3 | 13 ± 4 |
| 7 | 1840 ± 5 | 18 ± 7 | 5.5 | 7 ± 2 |

*The first column* contains the experimental values of the resonance masses and their errors (obtained by fitting procedure).

*The second column* contains the experimental values of the total width of the resonances (obtained by fitting procedure).

*The third column* contains the statistical significances of the resonances, determined by formula: S.D. = $N_{Res.}/\sqrt{N_{back.}}$, where $N_{back}$ – the number of background events into a resonance area.

*The fourth column* contains the resonance cross-sections with their errors. For the cross section values and errors, we have taken into account the cross section for the reaction $np \rightarrow np\pi^+\pi^-$ at $P_n = 5.20\, GeV/c$ ($\sigma_{np \rightarrow np\pi^+\pi^-} = (6.22 \pm 0.28)\, mb$) [10].

**4. Rotation Bands**

All the system of peculiarities which is observed by us can be described with two rotational bands on dependency from the orbital momentum L:

$$M_L = M_0 + KL(L+1), \qquad (1)$$

where:  L – value of orbital momentum
  $M_0$ – starting point of rotation band;
  $M_L$ – calculated value of mass.
  Values of the *K* coefficients are fitted from the experiment.

**K = 16.8 MeV/c²**            Table 2

| L | $M_L$, MeV/c² | $M_{exp} \pm \Delta M_{exp}$, MeV/c² |
|---|---|---|
| 0 | 1503 | 1501 ± 5 |
| 1 | 1537 | 1538 ± 3 |
| 2 | 1604 | 1598 ± 4 |
| 3 | 1705 | 1705 ± 4 |
| 4 | 1839 | 1840 ± 5 |

**K = 11.4 MeV/c²**            Table 3

| L | $M_L$, MeV/c² | $M_{exp} \pm \Delta M_{exp}$, MeV/c² |
|---|---|---|
| 0 | 1433 |  |
| 1 | 1456 | bump |
| 2 | 1501 | 1501 ± 5 |
| 3 | 1570 | bump |
| 4 | 1661 | 1665 ± 4 |
| 5 | 1775 | 1775 ± 4 |

Tables 2, 3 contain:
  *The first column* - the values of the orbital momentum;
  *The second column* - the resonance masses, calculated by formula (1) with corresponding values of the *K* coefficient;
  *The third column* - the experimental values of the resonance masses (see Table 1).

Our determination of the rotation bands does not contradict to the fact that there can be placed two very narrow resonances, which give as a result the average value of an experimental mass equal to M=1501 MeV/c².

## 5. Spin Estimation

We have tried to estimate the values of spins of the observed resonances in $nK^+$- system.

To do this, we constructed the distributions of emission angles of neutrons from the resonance decay with respect to the direction of resonance emission in the general c.m.s. of the reaction. All the values are transformed to the resonance rest system (helicity coordinate system). The backgrounds are constructed using events at the left and at the right of the corresponding resonance band and subtracted using the weight in proportion to a contribution of a background into resonance region. The result distributions (named "Cos Θ" distribution) are described by a sum of even-power Legendre polynomial (when approximating it was essential that errors in coefficients of the selected Legendre polynomials should be less then 50%) with maximum power being equal to (2J-1), where J is a spin of the resonance (for half-integer spins) In such a manner the value of the lower limit of the resonance spin is estimated The authors are grateful to Dr. V. L. Lyuboshitz for the arrangement of the corresponding formulas.

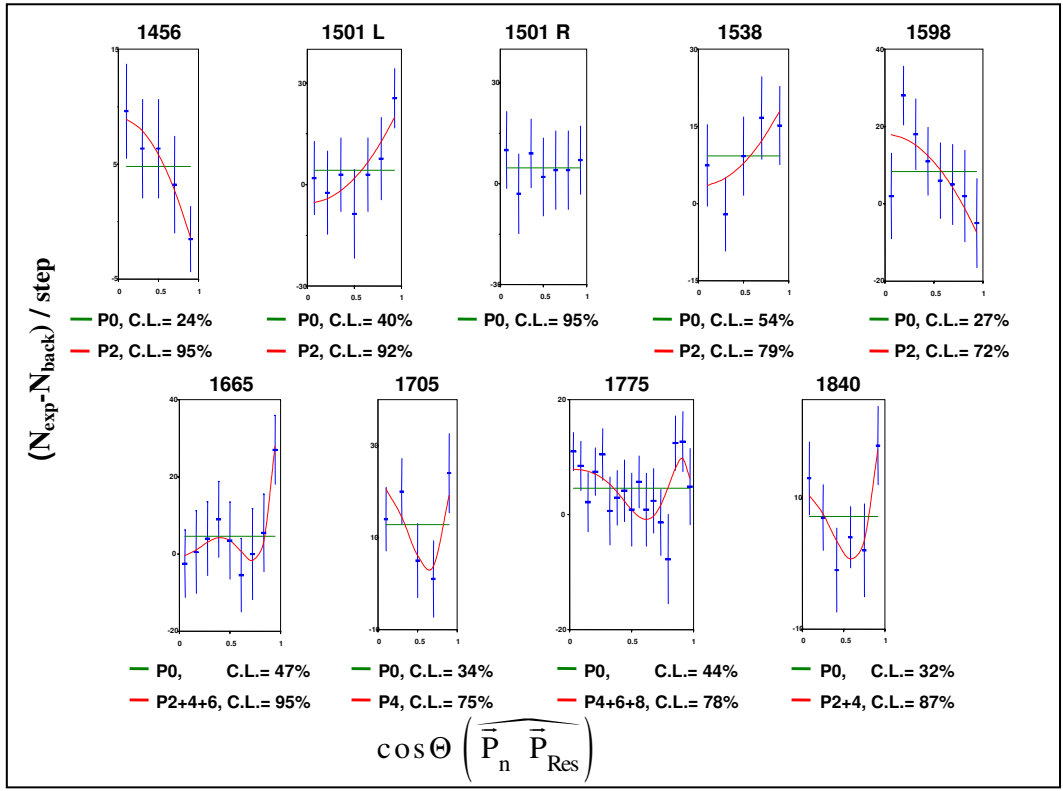

Figure 3

The Cos Θ distributions for the resonance spin estimations.
Masses under investigation and statistical significances of description of corresponding distributions are shown near every graph.

The figures noted as "1501L" and "1501R" show the corresponding distributions for the narrow peaks at the masses 1498 MeV/c$^2$ and 1507 MeV/c$^2$ which have been got due to inspection to mass region ≈1500 MeV/c$^2$ with small binning and which give as a result the average value of an experimental mass equal to M=1501 MeV/c$^2$.

## 6. Results

We calculate the parity of the observed resonances, using the formula: $P_{nk} = P_n * P_k * (-1)^L$, where L - value of orbital momentum (see Tab. 2, 3).

The mass resolution function [see 10] grows with increasing mass as:

$$\Gamma_{res}(M) = 4.2\left[\left(M - \sum_{i=1}^{2} m_i\right)/0.1\right] + 2.8, \text{ where:}$$

M – the mass of the resonance, $m_i$ – the rest mass of the particles composing this resonance,
M and $m_i$ are in GeV/c$^2$; coefficients 4.2 and 2.8 are in MeV/c$^2$.

For example, the value of the mass resolution is equal to ≈7 MeV/c² for the resonance at the mass of $M_{Res}$= 1538 MeV/c²

The true width of a resonance is obtained by formula: $\Gamma_{Res}^{true} = \sqrt{\left(\Gamma_{Res}^{exp}\right)^2 - \left(\Gamma_{res}\right)^2}$.

The results of the present work are in Table 4.

Table 4

|   | $M_{Res} \pm \Delta M_{Res}$, MeV/c² | $\Gamma_{Res}^{true} \pm \Delta\Gamma_{Res}^{true}$, MeV/c² | $J^P$ |
|---|---|---|---|
| 1 | 1501 ± 5 | 12 ± 8 | 3/2⁻ 1/2⁻ |
| 2 | 1538 ± 3 | 5 ± 5 | 3/2⁺ |
| 3 | 1598 ± 4 | 9 ± 5 | 3/2⁻ |
| 4 | 1665 ± 4 | 10 ± 7 | 7/2⁻ |
| 5 | 1705 ± 4 | 13 ± 7 | 5/2⁺ |
| 6 | 1775 ± 4 | 23 ± 9 | 9/2⁺ |
| 7 | 1840 ± 5 | 25 ± 9 | 5/2⁻ |

*The first column* contains the experimental values of the resonance masses and their errors.

*The second column* contains the true values of the total width of the resonances.

*The third column* contains the values of the spin-parity of resonances (the lower limit of spins are presented).

*Acknowledgements*

The authors are grateful to prof. A.I. Malakhov and prof. S. Vokal for the equipment and organize help to the processing of works at the *np*-collision investigations at the 1m HBC of the VBLHE and for help to the presentation of obtained results.

We also thank to prof. V.L. Lyuboshitz for the persistent help during discussions on physical results.

Our peculiar grateful to the group of operators of the chamber frames processing (the NEOFI subdivision of the VBLHE) for their hard and very professional work in the track preview and measurement.

Also the preview and measuring devices service group headed by V.F. Rubtzov (the LIT of the JINR) is worthy of big grateful.

**References**
1. D.Diakonov, V.Petrov, M.Polyakov. Z.Phys. A359, 305-314 (1997).
2. D.Diakonov Acta Physica Polonica B, No.1-2, Vol.25 (1994).
3. http://www.rcnp.osaka-u.ac.jp/~hyodo/research/Thetapub.html
4. R.Jaffe, F.Wilczek, hep-ph/0307341.
5. S.Capstick, P.R.Page, W.Roberts, hep-ph/0307019.
6. P.Zh.Aslanyan et.al., hep-ex/0403044.
7. Yu.A.Troyan e t al., Physics of Particles and Nuclei, Letters, vol.2No.1,2005,pp.20-28, Proceeding of the Romanian Academy, Series A, Vol.5, No.3/2004; hep-ex0404003.
8. A.P. Gasparian et al. JINR, 1-9111, Dubna, 1975, Pribory i Teknika Eksp., 1977, v.2, p.37.
9. Moroz V.I.et al. – Yad. Fiz., 1969, v.9, p.565
10. Besliu C.et al. - Yad. Fiz., 1986, v.63, p.888.
11. Yu.A.Troyan et al., Phys.Atom.Nuc., Vol.63, No.9, 2000, pp.1562-1573 [Yad.Fiz., vol.63, No.9. 2000, pp.1648-1659].
12. Troyan Yu.A. et al. – JINR Rapid Communication, 1996, №6[80] - 96, p.73.
13. Baldin A.M. et al., - Kinematics of Nuclear Reactions (Oxford Univ.Press, London, 1961, transl. of 1st Russ.ed; Atomizdat. M., 1968, 2nd ed.)